\documentclass[twocolumn,showpacs,preprintnumbers,amsmath,amssymb]{revtex4}
\usepackage{graphicx}% Include figure files
\usepackage{dcolumn}% Align table columns on decimal point
\usepackage{bm}% bold math

\begin{document}

\title{Velocity of sound in relativistic heavy-ion collisions}

\author{Bedangadas Mohanty and Jan-e Alam}

\medskip

\affiliation{Variable Energy Cyclotron Centre, Calcutta
   700064, India}

\date{\today}

\begin{abstract}

We have studied the rapidity distribution of secondary hadrons
produced in nucleus-nucleus collisions at ultra-relativistic 
energies within the ambit of the Landau's hydrodynamical 
model. A reasonable description of the data can also be
obtained by using the Bjorken's hydrodynamical model if
the boost invariance is restricted to a finite rapidity
range. The sensitivity of the hadronic spectra on the equation 
of state vis- a -vis the velocity of sound has been discussed.
The correlation between the velocity of sound and the freeze-out
temperature has been indicated.
The effects of the non-zero widths of various mesonic and baryonic
degrees of freedom up to the mass value $\sim 2.5$ GeV  is seen to
be small. 
\end{abstract}

\maketitle

\section{INTRODUCTION}

One of the important problem in the field of relativistic
heavy-ion collisions is to find out the  equation of state (EOS) of 
the matter formed after nuclear collisions at ultra-relativistic energies
~\cite{qm01}. 
Under the assumption of local thermal equilibrium, the EOS is the functional 
relation between pressure (P) and the energy density ($\epsilon$), where 
P and $\epsilon$ are related through the velocity of sound, $c_s$ which
is defined as
$c_{s}^{2} = ({\partial P} /{\partial \epsilon})_{isentropic}$~\cite{ll}. 
For a massless, non-interacting gas, $c_{s}^{2} = 1/3$
(ideal gas limit).  The velocity of sound plays a 
crucial role in the hydro-dynamical evolution of the matter created in
heavy-ion collision and affects, among others,
the momentum distribution of the particles
originating from the fluid elements at the freeze-out stage. 
In the present work we will assume
the order of the phase transition from QGP to hadrons to be first order.
However, it should be mentioned here that this issue is not  
fully settled yet. The phase transition from QGP to hadrons could
be weak first order, second order or it may be just a cross over
depending on the mass of the dynamical quarks~\cite{karsch}.
Consider a situation where quark-gluon plasma (QGP) is formed at the 
initial state. In such a scenario, the matter evolves from an initial QGP 
state to the hadronic phase via an intermediate mixed phase of QGP and 
hadrons due to the expansion of the system (hence cooling) in a first 
order phase transition scenario.  
Finally the system disassembles to 
hadrons (mainly pions) at the freeze-out where the interaction among the 
particles become too weak to maintain the equilibrium. The velocity of sound 
is  very different in the three stages of expansion mentioned above 
reflecting the interaction among the constituents  of the matter 
in the three stages. While in the QGP phase, it should in principle 
approach the ideal gas limit, in the mixed phase it should reduce to zero 
due to vanishing pressure gradient, indicating ``softness'' of the EOS.
Then below the critical temperature, it should have a value that reflects the 
presence of interacting hadrons in the system. 

Relativistic hydro-dynamical models have been routinely used to describe
the multiplicity distribution in the rapidity space, transverse mass spectra 
of hadrons etc produced in nuclear collisions. 
One of the aim of the present
work is to calculate the velocity of sound for hadronic system and
compare the results with lattice QCD calculations. In the present
calculation we have taken into account the non-zero width of the
unstable hadrons. We study 
the rapidity distribution of the particles within the 
framework of relativistic hydro-dynamical models proposed by
Landau and collaborators~\cite{ldl} and by Bjorken~\cite{jdb}.  
The results of the analysis performed within these models 
will be compared with  experiments to
extract the velocity of sound.
It is expected that at AGS  
and also at SPS energies, the Landau hydro-dynamical model 
can be applied although the Bjorken hydro-dynamical model
has been used at SPS energies. The main criticisms of Landau model
are: (i) neglect of leading particle effects and (ii) removal of 
radiation energy due to the deceleration required in this model for
full stopping. These difficulties, however, can be removed 
if one assume that during the collisions the valence quarks move 
without much interaction 
and the energy carried by the gluon fields is stopped in 
the collision volume~\cite{pc}. This assumption is justified because
the gluon-gluon interaction cross-section is larger
than quark-quark interaction due to larger color degeneracy of
the gluons. The gluon field thermalizes after a time 
$\tau_i$, providing the initial condition of the Landau
model. In this picture  the removal of energy of the decelerated gluons
fields due to the bremsstrahlung is prohibited by 
the color confinement mechanism. Under these conditions, it is obvious
that only a fraction of the beam energy is stopped in the collisions,
which can be taken into account by introducing an in-elasticity factor
in the model~\cite{pc}. 

Bjorken hydrodynamical 
model predicts a plateau structure for the rapidity distribution of the
fluid elements which is not observed experimentally for
the entire rapidity range. It has been indicated in~\cite{eskola} from
perturbative QCD that the initial energy distribution is a broad
Gaussian in rapidity even at LHC energies. However, we will show that 
a good description
of the data is possible at AGS and SPS energies in the framework of
the Bjorken's model if we treat the upper limit of the fluid rapidity
as a parameter. In other words the boost invariance has to be limited
in a finite rapidity range for the description of the data.

The paper is organized as follows. In the next section we obtain the velocity
of sound for a hadronic model and compare the result with the lattice QCD 
calculations~\cite{karsch}
and those obtained from a simple confinement model 
as discussed in Ref~\cite{rasww}
. In section III 
we extract the velocity of sound from the particle number density 
in rapidity space at AGS and SPS energies.
Finally in section IV we present summary and discussions. 

\section{Equation of State for hadron gas}

\begin{table}
\caption{ Particles taken for calculation.
\label{table1}}
\begin{tabular}{lcccr}
\tableline
Baryons&Mesons\\
\tableline
p& $\pi^{+,-,0}$\\
n& $\eta$(547-1440), $\eta^{\prime}$ \\
N(1440 - 2600) & $f_{0}$(800-1710),$f_{1}(1285,1420)$,$f_{2}$(1270-2340) 
\\
$\Delta$(1232 - 2420) & $\rho$(770-1700),$\rho_{3}$\\
$\Lambda$(1115 - 2350) &$\omega$(782-1650),$\omega_{3}$\\
$\Sigma^{+,-,0}$ & $a_{0}$(980-1450), $a_{1}$ \\
$\Sigma$(1382 - 1820) & $\phi$,$\phi_{3}$,$\pi$(1300-1800),$\pi_{2}$ \\
$\Xi^{0,+}$ & $f_{4}$,$h_{1}$,$K_{1}$(1270-1400)\\
$\Xi$(1530 - 1820) & $a_{2}$,$a_{4}$,$b_{1}$,$K_{2}$ \\
$\Omega^{-}$ & $K^{\pm,0}$, $K^{0}_{L,S}$,$K^{*}$,$K^{*}_{0,2,3,4}$ \\
 
\tableline
\end{tabular}
\end{table}

For the study of the EOS for the hadronic gas we take all the hadrons 
as listed in the particle data book up to the strange sector. 
The complete list is given in Table~I. 
The thermodynamical quantities like energy density ($\epsilon$), and 
pressure ($P$) can be calculated using the standard relations,
\begin{eqnarray}
\noindent\epsilon  = \sum_{hadrons}\frac{g}{(2 \pi)^{3}} 
\int {E(\vec{p}) f(E)
\rho(M)\,dM^2\,d^{3}p} \nonumber \\
P  = \sum_{hadrons}\frac{g}{(2 \pi)^{3}} 
\int {\frac{{\vec{p}}^{2}}{3E} f(\vec{p})
\rho(M)\,dM^2d^{3}p} 
\label{eq1}
\end{eqnarray}
where $E = \sqrt{{\vec{p}}^{2} + M^{2}}$ is the energy of the particle
of three momentum $\vec{p}$ and  invariant mass $M$,
$g$ is the internal degrees of freedom and 
$f(E)$ is the well-known  thermal distribution for bosons and fermions
which is given by $f(\vec{p}) = \Bigl[ exp((E-\mu)/T) \pm 1 \Bigr]^{-1},$ 
where $\mu$ is the chemical potential, $T$ is the temperature.  
The summation in Eq.~(\ref{eq1}) is carried out for all
the hadrons up to strange sector~\cite{pdg}. 
The sensitivity of the EOS on the hadronic and electromagnetic spectra
has been studied in ~\cite{soll}, with less number of hadrons and the
width of the hadrons are ignored.
$\rho(M)$ denotes the spectral function of the hadrons of pole mass $m$ 
and width $\Gamma$ given by,
\begin{equation}
\rho(M)=\frac{1}{\pi}\,\frac{M\Gamma(M)}{(M^2-m^2)^2+M^2\Gamma^2}
\end{equation}
Note that for stable
particle $\rho(M)\,\rightarrow\,\delta(M^2-m^2)$ as $\Gamma\,\rightarrow\,0$. 
To evaluate the spectral function of a hadron in a thermal
bath one should consider all the elastic and in-elastic 
processes through which the hadrons under consideration interact
with all the constituents of the thermal bath~\cite{weldon}.
These interactions give rise to the momentum dependent effective mass
and widths. However, in the present work we restrict to the
vacuum values of these quantities which are taken from particle
data book~\cite{pdg}. The spectral function
can also be related to the two body phase shift ($\delta$) 
of the various processes as $\rho(M)\propto d\delta(M)/dM$,
leading to the results obtained by S-matrix formalism of the statistical
mechanics proposed by Dashen et al.~\cite{dashen}.

From Eq.~\ref{eq1} one
can calculate the entropy density ($s$) by using the relation,  
$s = (\epsilon + P - \mu n)/T$ where $n$ is the
net baryon density given by,
\begin{equation}
\noindent n  = \sum_{baryons}\frac{g}{(2 \pi)^3} 
\int f(E) \rho(M)\,dM^2\,d^3p \nonumber \\
\end{equation}
$g_{eff}$  is the effective degeneracy which is parameterized 
as $g_{eff}(\mu,T) = 90 s/(4 \pi^2 T^3)$. 
The velocity of sound for $\mu=0$ can be calculated from 
the $g_{eff}$ as,
\begin{equation}
c_{s}^{-2} =\frac{T}{s}\frac{ds}{dT}=3 + \frac{T}{g_{eff}}\frac{dg_{eff}}{dT}
\label{eq2}
\end{equation}

We will  assume the chemical potential to be zero for mesons and 
consider two cases for baryonic chemical potential,  $\mu = 0$ and 
$200$ MeV. 

%%%%%%%%%%%%%%%%%%%%%%%%%%%%%%%%%%%%%%%%%%%%%%%%%%%%%

\begin{figure}
\begin{center}
\includegraphics[scale=0.5]{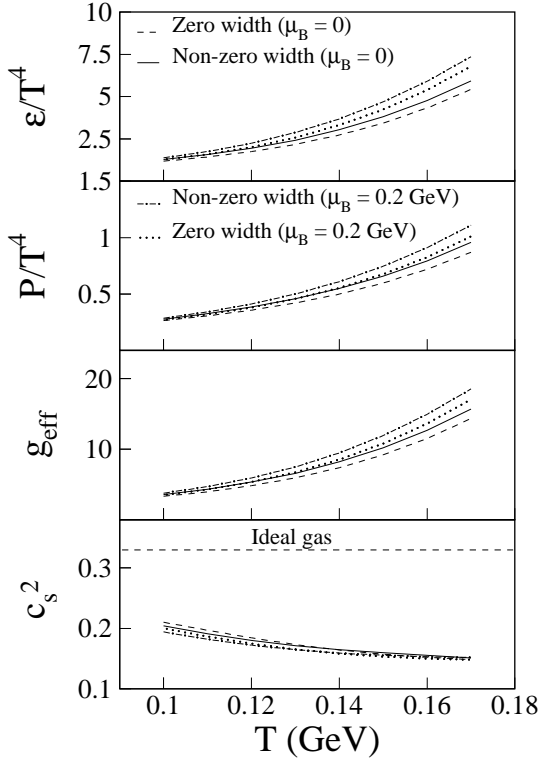}
\caption{ Variation of energy density ($\epsilon$), 
pressure ($P$), effective statistical degeneracy ($g_{eff}$)
and velocity of sound ($c_s$) with temperature ($T$).
}
\label{fig1}
\end{center}
\end{figure}

In  Fig.\ref{fig1} we plot the variation of various thermodynamic
quantities: $\epsilon$ , $P$, $g_{eff}$ and $c_{s}^{2}$ 
as a function of temperature. 
We make two observations: (a) the effects of the finite width in the masses
of the hadrons on the thermodynamical quantities turn out to be small
and (b) the effect of the finite  baryonic chemical potential leads to
a slightly larger values of  $\epsilon$ , $P$ and $g_{eff}$ but lower
values of $c_{s}^{2}$ at (typical) freeze-out 
temperature of 120 MeV. 

For comparison of our results with those from the lattice QCD 
calculations~\cite{karsch}, 
we parameterize the variation of energy density ($\epsilon$)
of lattice QCD results with temperature as follows~\cite{bm}:  
\begin{equation}
\epsilon = T^{4} A~tanh\left(B (\frac{T}{T_{c}})^C\right),
\label{eq3}
\end{equation}
where $A, B$ and $C$ are parameters whose values are 12.44, 0.517
and 10.04 respectively. Note that the effect of baryons in the EOS
is neglected here.

The evolution of the system 
under the assumption of boost-invariance along the longitudinal
direction 
is governed by the equation~\cite{jdb},
\begin{equation}
\frac{d \epsilon}{d \tau}  + \frac { \epsilon + P}{\tau} = 0 ; P = c_{s}^2 \epsilon
\label{eq4}
\end{equation}
where $c_{s}$ is the velocity of the sound in the medium.
We would like to mention here that the velocity of sound
sets the expansion time scale ($(\tau_{exp})^{-1}\sim\,(1/\epsilon)
d\epsilon/d\tau=(1+c_s^2)/\tau$) for the system. This time scale
should be larger than the collision time scale 
($(\tau_{coll})^{-1}\sim\,n\sigma v$,
where $\sigma$ is the cross section, $v$ is velocity and $n$ is
density) for thermal equilibrium to be maintained in the system. Therefore,
determination of velocity of sound becomes very important for the study
of the space time evolution of the system in general and 
hadronic spectra in particular.

We obtain the expression for the velocity of sound as, 
(see also~\cite{bm}),
\begin{equation}
c_{s}^2 = \Bigl[ 3 + BC (\frac{T}{T_c})^C \frac{1}{cosh(B (\frac{T}{T_c})^C)~ sinh(B (\frac{T}{T_c})^C)} \Bigr]^{-1}
\label{eq5}
\end{equation}

The results for the confinement model were obtained from~\cite{rasww}
by extracting  $g_{eff}(T)$  by using the relation 
$\epsilon/T^4=(\pi^2/30)g_{eff}$ and  
Eq.~\ref{eq2}.

\begin{figure}
\begin{center}
\includegraphics[scale=0.5]{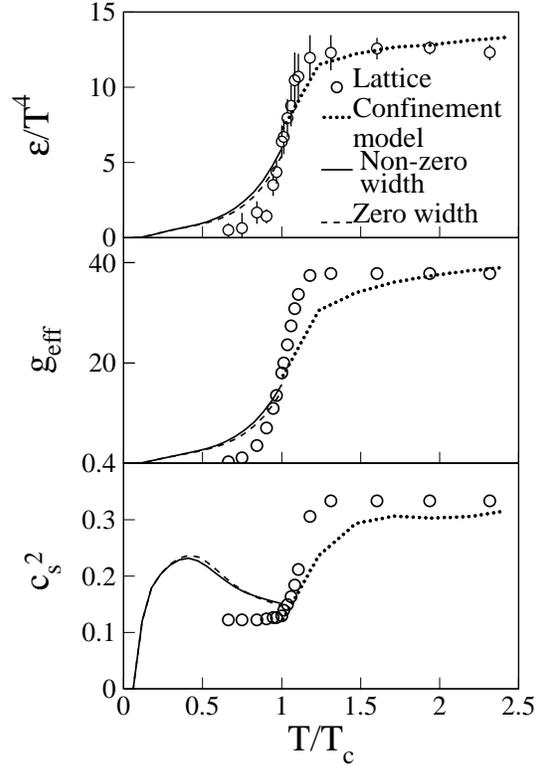}
\caption{Variation of energy density, $g_{eff}$ and velocity of sound using 
the combined results of hadron gas, confinement model  and lattice 
calculations with temperature. The lattice results for energy density
has been read from \protect{\cite{karsch}}. $g_{eff}$ and $c_s^2$
are derived by using the equations mentioned in the text.
}
\label{fig2}
\end{center}
\end{figure}

%%%%%%%%%%%%%%%%%%%%%%%%%%%%%%%%%%%%%%%%%%%%%%%%%%%%%

In Fig.\ref{fig2} the energy density, effective statistical 
degeneracy and the velocity of sound evaluated in the hadronic
model at zero baryonic chemical potential  
are compared
with the lattice data
($\mu$ = 0) and those obtained form the confinement model. 
We observe that, 
(a) our results match with those from the lattice
and the confinement model near $T_{c}$, 
(b) they are higher than those obtained from lattice below $T_{c}$ (In 
this context we would like to point out two things. Firstly, lattice 
results usually employ quark masses which are too large,
leading to larger pion masses, $m_{\pi}^{lat} \ge
3m_{\pi}^{physical}$. The resulting thermal suppression of these
degree of freedom causes a considerable discrepancy for lattice EOS 
with respect to EOS for hadron gas. Secondly, the lattice calculations 
below $T_{c}$ has large errors),
(c) at temperatures below $T_{c}$ the velocity of sound shows an 
interesting trend, it increases with decrease in temperature then
falls to zero as temperature of the system approaches zero and
(d) the complete EOS for the system with initial QGP state, may be
those obtained from lattice for temperatures above $T_{c}$ and those
given by the present calculations for temperatures below $T_{c}$.

The velocity of sound, which is an input to the hydro-dynamical evolution of
the system via EOS influence the rapidity distribution of the hadrons. 
In the next section we try to extract the velocity of sound from the 
rapidity distributions of various hadrons produced from heavy-ion 
collisions for various collision energies using hydro-dynamical model. 
We then compare the values obtained with the results of the model 
under consideration.

\section{Rapidity distribution of secondaries  at AGS and SPS energies}
There are two well known hydro-dynamical models as mentioned 
before~\cite{ldl,jdb} for the description of the  
space time evolution of the system formed after the collision of heavy-ions. 
It may be mentioned that in the coherent interactions the collective 
effect is an important feature, unlike incoherent collisions where 
collision is considered to be a succession of independent nucleon-nucleon 
interactions.  The basic difference is, in Landau's model, in the 
center of mass of the collision two nuclei would be stopped and all the 
kinetic energy would be used up for the particle production, while  
the baryon transparency in the mid-rapidity region 
is the basic feature of the Bjorken's model. However, 
the Landau's model can be applied to describe the system formed
after nuclear collision with appropriate initial conditions as
mentioned in the introduction. 
It is expected that the Landau hydrodynamics will work well 
at AGS and may be at SPS energies, while the Brojken hydrodynamics should 
in principle work well for energies at RHIC and LHC.

In the following we briefly discuss the rapidity
distribution of the particles produced in relativistic heavy-ion collisions
in Landau's  and Brojken's model. 

The amount of entropy ($dS$) contained within a (fluid) rapidity $dy$ in
the Landau hydro-dynamical model is given by~\cite{ldl,ouran},
\begin{equation}
\frac{dS}{dy} = - \pi R^{2}l s_{0} \beta c_{s} 
exp[{\beta \omega_{f}}] 
\Bigl[I_{0}(q) - \frac{\beta \omega_{f}}{q}I_{1}(q) \Bigr] 
\label{eq6}
\end{equation}
where, $q = \sqrt{\omega_f^{2} - c_{s}^{2}y^{2}}$, 
$\omega_{f} = ln(T_f/T_0)$, $T_f$ is the freeze-out 
temperature, $y$ is the rapidity, 
$R$ is the radius of the nuclei, $2l$ is the initial length, 
$s_{0}$ is the initial entropy density, 
$2\beta = (1-c_{s}^{2})/c_{s}^{2}$ and  $I_{0}$, $I_{1}$ are 
the Bessel's function. The quantity $\pi R^2l s_0$ is fixed
to normalize the experimental data at the mid rapidity. 
For $\omega_f>>c_sy$ the quantity $dS/dy$ can be approximated 
by a Gaussian distribution,
\begin{equation}
\frac{dS}{dy}\,\sim\,Const.\frac{\exp(-\frac{y^2}{2\sigma^2})}
{\sqrt{2\pi\sigma^2}} 
\label{eq6prime}
\end{equation}
where $\sigma=2\omega_f/(1-c_s^2)$.

The entropy density in the Brojken hydrodynamics is given by
~\cite{ouran},
\begin{equation}
\frac{dS}{dy} = \frac{A \pi R^{2}}{T_{f}} s_{f} exp[{-2\beta \omega_{f}}] 
\label{eq7}
\end{equation}
where, $A$ is a constant and $s_{f}$ is the entropy density at 
the freeze-out. It is to be noted that the quantity $dS/dy$ 
is independent of the rapidity in accordance with the assumption
of boost invariance along the longitudinal direction~\cite{jdb}. 

The multiplicity distribution of the secondaries is obtained by
folding the multiplicity density in rapidity space  mentioned above 
by the thermal  distribution of the fluid elements,

\begin{table}
\caption{ AGS results
\label{table2}}
\begin{tabular}{lcccr}
\tableline
Type&$c_{s}^{2}$&$\chi^{2}$&Probability\\
\tableline
proton 4 AGeV&1/5 (1/3)& 0.65 (34.5) & 1.0 (0.023)\\
proton 6 AGeV&1/5 (1/3)& 2.15 (40.2) & 1.0 (0.010)\\
proton 8 AGeV&1/5 (1/3)& 7.1 (74.8) & 0.999 (0.00001)\\
proton 11.6 AGeV&1/5 (1/3)& 1.49 (31.1) & 1.0 (0.0388)\\
\tableline
\end{tabular}
\end{table}

\begin{equation}
\frac{dN}{dY} = \frac{g}{(2\pi)^{3}} 
\int {\frac{dN}{dy} f(M_{T},y,Y) M_{T} cosh(Y-y) dy d^{2}p_{T}} 
\label{eq8}
\end{equation}
where, $dN$ is the number of particles within the rapidity 
interval $dY$,
$M_T$ is the transverse mass ($=\sqrt{p_T^2+M^2}$) of the 
particle, $E$ = $M_{T} cosh(Y-y)$, is the energy and 
$f(E)$ = $1/(exp[E/T] \pm 1$). $dS/dy$ is related to $dN/dy$
by a constant factor (see~\cite{bm} for details). Performing 
the $p_T$ integration in the above equation we obtain,
\begin{equation}
\frac{dN}{dY} \propto 
T^3 \int \frac{dN}{dy}h(y,Y;m,T)
\exp(-mcosh(Y-y)/T)\,dy  
\label{eq11}
\end{equation}
where
\begin{equation}
h(y,Y;m,T)=\frac{m^2}{T^2}+2\frac{m}{T}\frac{1}{cosh(Y-y)}
+\frac{2}{cosh^2(Y-y)}
\label{eq12}
\end{equation}
It may noted from Eqs.\ref{eq11} and \ref{eq12}  that the particle mass 
tend to make the rapidity distribution narrower.

It may be mentioned that, while in Landau's model the width of the 
rapidity distribution is sensitive to the velocity of sound and the 
freeze-out temperature, in the case of Brojken's model it is not
(because of its independence on rapidity, the quantity,
dS/dy can be fixed
here by the normalization at the mid-rapidity). While integrating
over $y$ in Eq.~\ref{eq11} we treat the range of $y$ as a parameter
in case of Bjorken's hydrodynamics~\cite{es}. In case of Landau's
hydrodynamics the integration limit for rapidity is infinite.
We apply the Landau's model to 
the rapidity distributions of the produced hadrons at AGS and SPS 
energies. It will be seen later that the Landau model fits the pion
rapidity spectra well at lower SPS energies, however at highest
SPS energy the description is not very satisfactory, 
therefore we do not attempt to apply the model to RHIC 
energies. We fix the freeze-out temperature $T_{f}$ to be 120 MeV 
a value obtained by studying the transverse momentum 
distributions of the hadrons~\cite{bkp,na49}.

%%%%%%%%%%%%%%%%%%%%%%%%%%%%%%%%%%%%%%%%%%
\begin{figure}
\begin{center}
\vskip -1.2 cm
\includegraphics[scale=0.4]{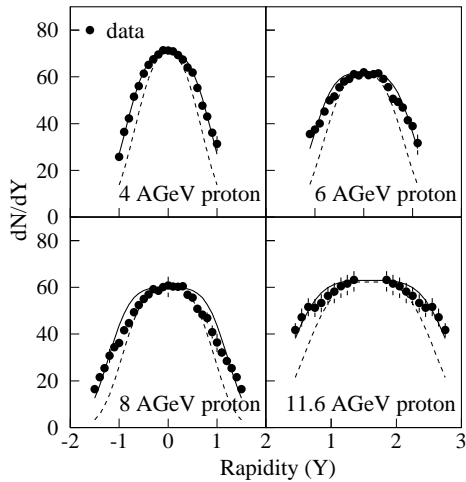}
\caption{ Rapidity spectra for protons at 4,6,8, and 11.6 AGeV Au+Au collisions
compared to rapidity spectra obtained from Landau hydrodynamics
with velocity of sound 0.2 (solid line) and 0.333 (dashed line).
}
\label{fig3}
\end{center}
\end{figure}

The rapidity distribution of the protons~\cite{jlk,la} at AGS energies are
well described by the Landau hydro-dynamical model with velocity of sound,
$c_s^2=1/5$, a value different from that corresponding to an ideal 
gas (Fig.\ref{fig3}). For $c_s^2=1/5$ the values of the chi-square  
is considerably smaller than for $c_s^2=1/3$ (Table~\ref{table2}).  
The peak of the distribution at the mid-rapidity in the experimental data 
seems to indicate a large deposition of collision energy  in these 
interactions, where the Landau's hydro-dynamical picture is applicable.

%%%%%%%%%%%%%%%%%%%%%%%%%%%%%%%%%%%%%%%%%%%%%%%%%%%%%
\begin{figure}
\begin{center}
\vskip -1.2 cm
\includegraphics[scale=0.4]{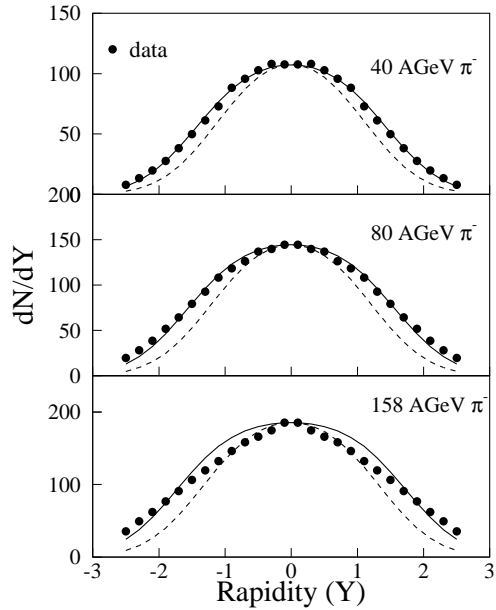}
\caption{ Rapidity spectra for pions at 40,80,158 A GeV Pb+Pb collisions
compared to rapidity spectra obtained from Landau hydrodynamics
with velocity of sound 0.2 (solid line) and 0.333 (dashed line).
}
\label{fig4}
\end{center}
\end{figure}

\begin{table}
\caption{ SPS results for pions
\label{table3}}
\begin{tabular}{lcccr}
\tableline
Type&$c_{s}^{2}$&$\chi^{2}$&Probability\\
\tableline
$\pi^{-}$ 40 AGeV&1/5 (1/3)& 1.03 (51.7) & 1.0 (0.0013)\\
$\pi^{-}$ 80 AGeV&1/5 (1/3)& 2.6 (66.3) & 0.999 (7 $\times~10^{-7}$)\\
$\pi^{-}$ 158 AGeV&1/5 (1/3)& 19.0 (146.0) & 0.80(4 $\times~10^{-19}$)\\
\tableline
\end{tabular}
\end{table}

\begin{table}
\caption{ SPS results for kaons
\label{table4}}
\begin{tabular}{lcccr}
\tableline
Type&$c_{s}^{2}$&$\chi^{2}$&Probability\\
\tableline
$K^{+}$ 40 AGeV&1/5 (1/3)& 0.463 (18.0) & 1.0 (0.38)\\
$K^{-}$ 40 AGeV&1/5 (1/3.3)& 0.56 (0.16) & 1.0 (1.0)\\
$K^{+}$ 80 AGeV&1/4.5 (1/3)& 0.4 (12.7) & 1.0 (0.75)\\
$K^{-}$ 80 AGeV&1/5 (1/3.3)& 3.0 (0.67) & 0.99 (1.0)\\
$K^{+}$ 158 AGeV&1/4.5 (1/3)& 0.39 (15.6) & 1.0 (0.55)\\
$K^{-}$ 158 AGeV&1/5 (1/3.5)& 3.8 (0.8) & 0.99 (1.0)\\
\tableline
\end{tabular}
\end{table}

\begin{figure}
\begin{center}
\vskip -1.2 cm
\includegraphics[scale=0.4]{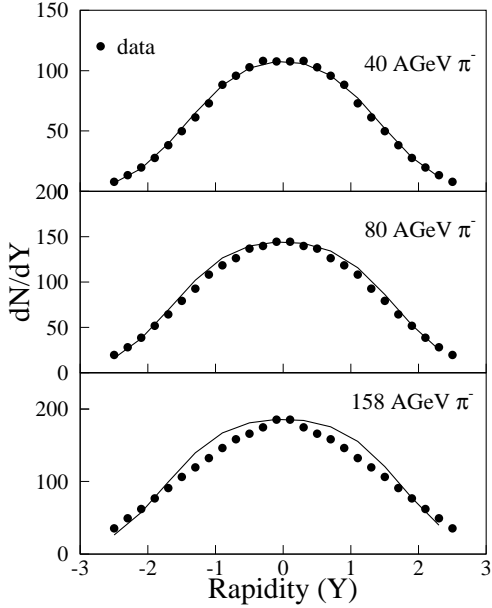}
\caption{Rapidity spectra for pions at 40,80,158 A GeV Pb+Pb collisions
compared to rapidity spectra obtained from Bjorken's hydrodynamics.
The value of $y_{max} (=-y_{min})$ used in the integration of 
Eq.~\ref{eq11} are 1.44, 1.65 and 1.6 for beam energies 40 AGeV,
80 AGeV and 158 AGeV respectively. 
}
\label{fig4a}
\end{center}
\end{figure}

For SPS energies the density of pions in rapidity space~\cite{mvl} 
is well reproduced (Fig.\ref{fig4}) within the ambit of Landau's model
with $c_s^2=1/5$ and $T_f=120$ MeV for 40 and 80 GeV/A beam energies.  
For $E_{beam} = 158$ GeV/A the $\chi^2$ worsens (Table~\ref{table3}). 
The Bjorken's model also
does not give a satisfactory description of the data at
higher SPS energies (158 AGeV). However, as shown in Fig.~\ref{fig4a}
at lower energies the Bjorken's model give a reasonable description of
the data.

\begin{figure}
\begin{center}
\vskip -1.2 cm
\includegraphics[scale=0.4]{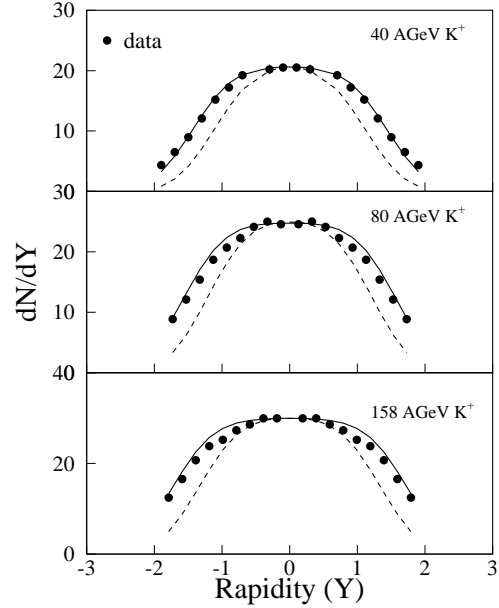}
\caption{Rapidity spectra for $K^{+}$ at 40,80,158 A GeV Pb+Pb collisions
compared to rapidity spectra obtained from Landau hydrodynamics
with velocity of sound 0.2 (solid line) and 0.333 (dashed line) and 
$T_{f}$ = 120 MeV.
}
\label{fig5}
\end{center}
\end{figure}

\begin{figure}
\begin{center}
\vskip -1.2 cm
\includegraphics[scale=0.4]{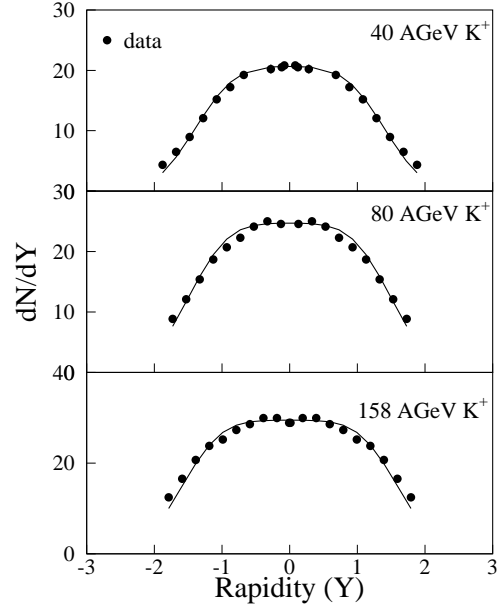}
\caption{Rapidity spectra for kaons at 40,80,158 A GeV Pb+Pb collisions
compared to rapidity spectra obtained from Bjorken's hydrodynamics.
The value of $y_{max} (=-y_{min})$ used in the integration of 
Eq.~\ref{eq11} are 1.4, 1.5 and 1.6 for beam energies 40 AGeV,
80 AGeV and 158 AGeV respectively. 
}
\label{fig5a}
\end{center}
\end{figure}

\begin{figure}
\begin{center}
\vskip -1.2 cm
\includegraphics[scale=0.4]{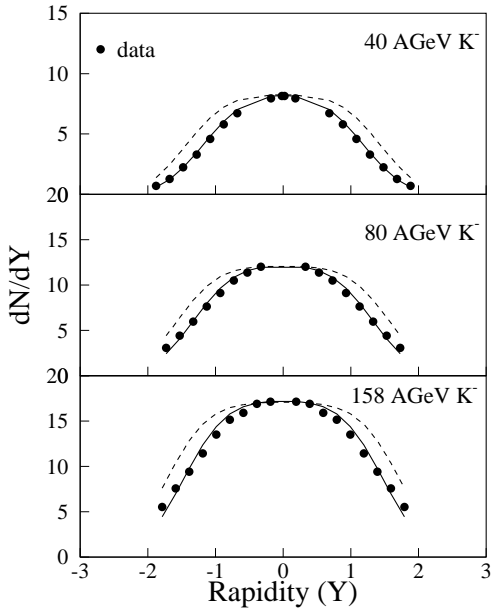}
\caption{ Rapidity spectra for $K^{-}$ at 40,80,158 A GeV Pb+Pb collisions
compared to rapidity spectra obtained from Landau hydrodynamics
with velocity of sound 0.2 for $T_{f}$ = 120 MeV (dashed line) and 132 
MeV (solid line).
}
\label{fig6}
\end{center}
\end{figure}

Interestingly, the K$^+$ spectra is well reproduced 
(Figs.\ref{fig5} and \ref{fig5a}) with the same freeze-out 
temperature and velocity of sound  as mentioned above for all the beam 
energies for both the models. 
However, the K$^-$ spectra (Fig.\ref{fig6}) is reproduced with a 
slightly higher freeze-out temperature, 132 MeV. A good description of
the data for 40 and 80 GeV/A beam energies is also obtained for
$T_f$=120 MeV and $c_s^2=1/3.5$($\chi^2=0.1$ and 0.3
respectively). For $E_{beam} = 158$ GeV/A
we obtain $c_s^2=1/4$ for same freeze-out temperature ($\chi^2=0.3$).
The values of $\chi^2$ are shown in Table~\ref{table4}.
It is well-known that the system formed in nuclear collisions
at AGS and SPS energies has non-zero baryonic density and in
such a situation the differences in the values of the freeze-out
parameters for $K^+$ and $K^-$ are expected (see~\cite{uh} for details).

%%%%%%%%%%%%%%%%%%%%%%%%%%%%%%%%%%%%%%%%%%%%%%%%%%%%%

%%%%%%%%%%%%%%%%%%%%%%%%%%%%%%%%%%%%%%%%%%%%%%%%%%%%%
\begin{figure}
\begin{center}
\vskip -1.2 cm
\includegraphics[scale=0.4]{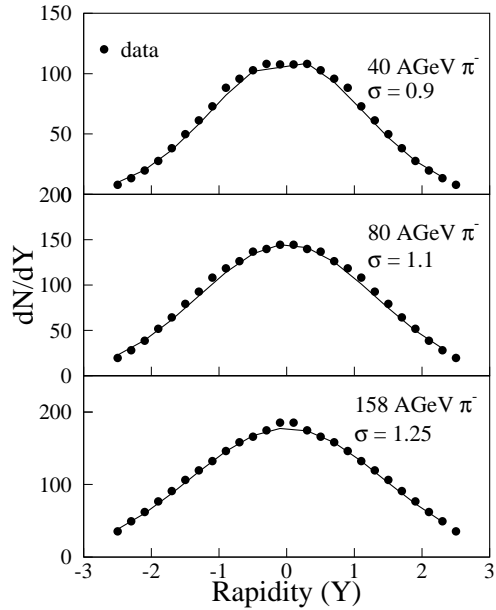}
\caption{Rapidity spectra for pions at 40,80,158 A GeV Pb+Pb collisions
compared to rapidity spectra obtained from Eqs.\ref{eq6prime} and
\ref{eq11}. The width of the Gaussian in Eq.~\ref{eq6prime} is a parameter 
here.
}
\label{pigauss}
\end{center}
\end{figure}

\begin{figure}
\begin{center}
\vskip -1.2 cm
\includegraphics[scale=0.4]{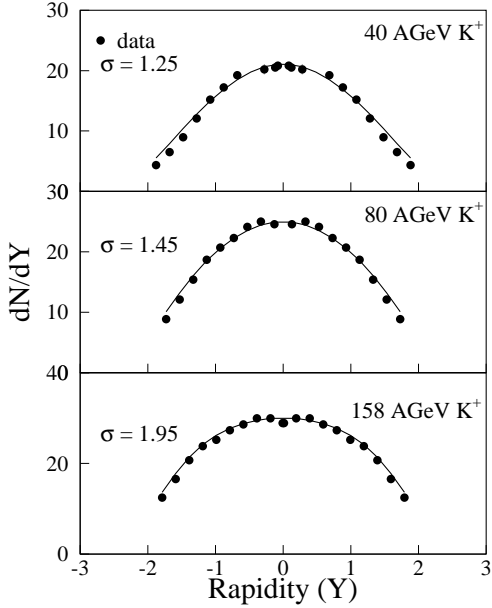}
\caption{ Same as ~\protect{\ref{pigauss}} for kaons.
}
\label{kaongauss}
\end{center}
\end{figure}

%%%%%%%%%%%%%%%%%%%%%%%%%%%%%%%%%%%%%%%%%%%%%%%%%%%%%
The pion data from  SPS at 40, 80 and 158 AGeV energies 
can also be reproduced if we use the Gaussian approximation of 
Eq. ~\ref{eq6prime} in Eq.~\ref{eq11} and treat the width of
the Gaussian ($\sigma$) as a parameter. Results are shown in Fig.\ref{pigauss}.
The value of $\sigma=0.9, 1.1$ and 1.25
for pion at beam energies 40, 80 and 158 AGeV respectively.
Results obtained for kaons in this procedure is shown
in Fig.~\ref{kaongauss}.  It is important to note that the
values of the widths of the Gaussian which represent the 
rapidity distribution of the fluid elements 
are 1.25, 1.45 and 1.95 for 40, 80 and 158 AGeV energies respectively
 and these values of $\sigma$ are consistently higher than the values 
obtained for
pions. The difference can be attributed to the mass difference
between pions and kaons because pions and kaons are  
subjected to the same longitudinal flow velocity.

%%%%%%%%%%%%%%%%%%%%%%%%%%%%%%%%%%%%%%%%%%%%%%%%%%%%%
%%%%%%%%%%%%%%%%%%%%%%%%%%%%%%%%%%%%%%%%%%%%%%%%%%%%%

\begin{figure}
\begin{center}
\vskip -1.2 cm
\includegraphics[scale=0.4]{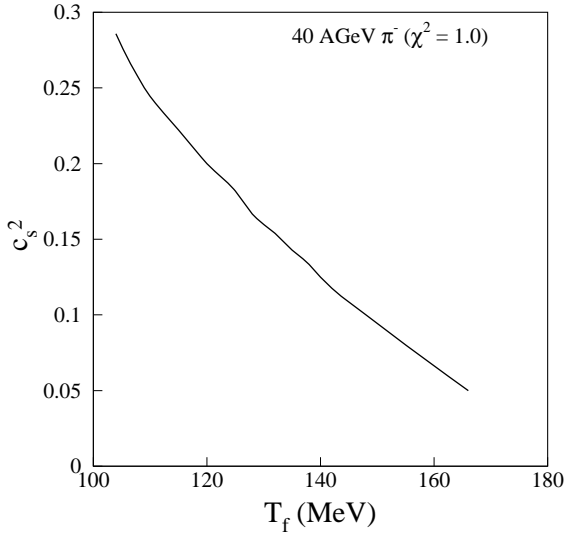}
\caption{The constant $\chi^2$ contour in $c_s^2-T_f$ plane for pions.
}
\label{chisq}
\end{center}
\end{figure}

In Fig.~\ref{chisq} we show the constant $\chi^2$ contour
in the $c_s^2-T_f$ plane. The results indicate that  we need
a value of $c_s^2$ less than 1/3 for a reasonable description
of the data.  
We observe that $c_{s}^{2}\,\sim\,1/5$  near the freeze-out 
for different collision energies and for various hadronic species.
This indicates a certain kind universality for the hadronic matter
produced in heavy-ion collisions at the late stage of the 
evolution (freeze-out). 
%%%%%%%%%%%%%%%%%%%%%%%%%%%%%%%%%%%%%%%%%%%%%%%%%%%%%

%%%%%%%%%%%%%%%%%%%%%%%%%%%%%%%%%%%%%%%%%%%%%%%%%%%%%
%%%%%%%%%%%%%%%%%%%%%%%%%%%%%%%%%%%%%%%%%%%%%%%%%555
%%%%%%%%%%%%%%%%%%%%%%%%%%%%%%%%%%%%%%%%%%%%%%%%%555
\section{Summary}
We have evaluated the velocity of sound in a hadronic model
with hadronic degrees of freedom up to strange sector.
This calculated velocity of sound is compared with the results
obtained from lattice QCD calculations~\cite{karsch}
and those obtained for the confinement model~\cite{rasww}.
The values compare well at the critical temperature.
The effects of non-zero width of hadrons on the EOS is found 
to be small.
Taking a non-zero baryonic chemical potential slightly decreases
the velocity of sound.
Fixing the value of the freeze-out temperature ($\sim 120$ MeV)
from the $p_T$ spectra of hadrons~\cite{bkp,na49} we find that
a value of $c_s^2=1/5$ give a good description of the data for Landau's
hydrodynamical model. The data is also well reproduced within
the ambit of the hydrodynamical model proposed
by Bjorken if the boost invariance is restricted to finite rapidity range.
The value of $c_s^2=1/5$ indicates that the expansion of the system
is slower in comparison to the ideal gas scenario ($c_s^2=1/3$) and
hence for such a system the maintenance of thermal equilibrium becomes
easier. We observe that different hadron species produced in
the nuclear collisions at different energies are well reproduced 
by a value of $c_s^2\sim 1/5$, indicating some kind of universality
of the matter at the freeze-out stage. 

Several other effects which are ignored in the present work
need to be mentioned at this juncture. For a complete description
of the space time evolution of the system formed after heavy ion 
collisions both  the transverse and longitudinal expansion should
be considered. The transverse momentum spectra of the hadrons 
is strongly affected by the transverse flow. The rapidity
distribution of the hadrons which is obtained after integrating 
out the transverse momentum may not be strongly affected
by the transverse flow. However, the normalization of the 
rapidity spectra may be substantially affected by the transverse
flow. The other quantities {\it e.g.} which may affect the normalization
are non-zero chemical potential of the mesons, arising from the lack
of chemical equilibrium in the system.  The hadrons (pions, kaons and
protons) originating for the decays of various mesonic ($\rho$,
$\omega$, $\phi$ etc) and baryonic ($\Delta$ etc) resonances  are ignored
here. If the distribution of the resonances is homogeneous 
in the rapidity then the effects of the hadrons originating from
their decays on the rapidity distributions  can be  ignored,
otherwise these effects may change the distribution depending
on the degree inhomogeneity and the abundances of the resonances at the 
freeze-out point. With the centrality of the collisions the normalization
and the width of the distribution of hadrons in rapidity space
changes. The normalization in the present work is treated as a parameter
and the change in the width with centrality is rather small. 
At 158 AGeV SPS energy the width of the rapidity distribution 
of charged particle increases from 1.5 to 1.62 when the centrality
changes from $(0-5)\%$ to $(25-35)\%$~\cite{na50}.

\acknowledgments{One of us (B.M.) is grateful to the Board of Research
on Nuclear Science and Department of Atomic Energy, 
Government of India for financial support in the form of Dr. K.S. Krishnan
fellowship. 
We would also like to thank Marco van Leeuwen of NA49 
Collaboration and Jennifer L. Klay of E895 Collaboration 
for providing the experimental data.
}

\end{document}